\documentclass[conference]{IEEEtran}
\IEEEoverridecommandlockouts
\usepackage{cite}
\usepackage{algorithm}
\usepackage{algorithmic}
\usepackage{graphicx}
\usepackage{textcomp}
\usepackage{xcolor}
\usepackage{graphicx}
\usepackage{subfigure}
\usepackage{subeqnarray}
\usepackage{bm}
\usepackage{amsfonts}
\usepackage{amsthm}
\usepackage{amssymb}
\usepackage{makecell}
\usepackage{calc}
\usepackage{mathrsfs}
\usepackage{flushend}
\usepackage{url}
\usepackage{caption}
\usepackage{subfloat}
\usepackage{color}
\usepackage{textcomp}
\usepackage{booktabs}
\usepackage{amsmath}
\usepackage[outdir=./]{epstopdf}

\ifCLASSOPTIONcompsoc
 \usepackage[caption=false,font=normalsize,labelfont=sf,textfont=sf]{subfig}
\else
 \usepackage[caption=false,font=footnotesize]{subfig}
\fi

\newtheorem{remark}{Remark}

\def\BibTeX{{\rm B\kern-.05em{\sc i\kern-.025em b}\kern-.08em
    T\kern-.1667em\lower.7ex\hbox{E}\kern-.125emX}}
\begin{document}

\title{Secure Dual-Functional Radar-Communication System via Exploiting Known Interference in the Presence of Clutter\\
{\large \textit {(Invited Paper)}}
}

\author{\IEEEauthorblockN{Nanchi Su, Zhongxiang Wei, Christos Masouros}
\IEEEauthorblockA{\textit{Department ofElectronic and Electrical Engineering} \\
\textit{University College London}, London, UK\\
\{nanchi.su.18, zhongxiang.wei, c.masouros\}@ucl.ac.uk}
}

\maketitle

\begin{abstract}
  This paper addresses the problem that designing the transmit waveform and receive beamformer aims to maximize the receive radar SINR for secure dual-functional radar-communication (DFRC) systems, where the undesired multi-user interference (MUI) is transformed to useful power. In this system, the DFRC base station (BS) serves communication users (CUs) and detects the target simultaneously, where the radar target is regarded to be malicious since it might eavesdrop the transmitted information from BS to CUs. Inspired by the constructive interference (CI) approach, the phases of received signals at CUs are rotated into the relaxed decision region, and the undesired MUI is designed to contribute in useful power. Then, the convex approximation method (SCA) is adopted to tackle the optimization problem. Finally, numerical results are given to validate the effectiveness of the proposed method, which shows that it is viable to ensure the communication data secure adopting the techniques that we propose.
\end{abstract}


\section{Introduction}
The spectrum resources is getting congested increasingly due to the rapid growth of wireless connections and mobile devices, which results in high auction price of the available wireless spectrum. As reported in \cite{Survey2018}, the Spanish government raised a total of \texteuro438 million for the sale of 5G frequencies and the government of South Korea paid \$3.3 billion for the spectrum bands at 3.5 GHz and 28 GHz in 5G network. While the spectrum bands allocated to radar is abundant, i.e., from 3-30 MHz band to 110-300 GHz band \cite{griffiths2014radar}. Since wireless communication has increasingly similar radio frequency (RF) front-end architectures, it allows radar share spectrum resources with communication cellular operators and improves the spectrum utilization efficiency, which motivates the development of dual-functional radar-communication (DFRC) \cite{hong2019ergodic}.
\\\indent In DFRC systems, radar and communication share the same hardware platform and
energy resources. The transmitted waveform is specifically designed as to serve for both purposes of target sensing and wireless communication, which implicates the possibility of information leakage. Intuitively, the radar beampattern is designed to concentrate the radiation power towards the direction of interest so as to improve the detection performance, in which case the target, as a potential eavesdropper, could readily surveil the information intended for communication users (CUs). To this end, physical layer (PHY) security is worthwhile being taken into consideration in security-critical DFRC designs.
\\\indent From the perspective of communication system, methods to secure the wireless communication systems are widely investigated in the past decades. As pioneered by Wyner \cite{wyner1975wire}, beamforming and precoder are designed to ensure the quality of service (QoS) at users of interest while limiting the signal strength of the potential eavesdroppers, which aims to yield an optimal difference in signal-to-interference-plus-noise ratio (SINR) at the intended users and the eavesdroppers, i.e., secrecy rate (SR) \cite{liu2014secrecy}. Furthermore, artificial noise (AN) is generated to further deteriorate the received signals at eavesdroppers \cite{su2020secure}. Additionally, Directional modulation (DM) has attracted more and more attention as an emerging approach to secure wireless communication systems in recent years. Unlike the SR based methods, DM technique adjusts the amplitude and phase of the symbols at the users of interest directly while scrambling the symbols in other undesired directions, which implies that the modulation happens at the antenna level instead of baseband level. By doing so, a constellation of certain modulation with low bit error rate (BER) could be received by desired users, while the constellation received by each eavesdropper will be distorted. On top of the research on DM, some studies focus on exploiting constructive interference (CI) \cite{masouros2015exploiting,liu2017efficient} through symbol-level precoding, which exploits known multiuser interference (MUI) as useful power by pushing the received signal away from the detection bound of the signal constellation..
\\\indent In this paper, the CI technique is investigated to enhance the PHY layer security. Specifically, we consider a DFRC BS which serves CUs while detecting a point-like target in the presence of clutter, where the target is treated as a potential eavesdropper which might intercept the information intended to CUs. In the problem formulation, the transmit waveform and the receive beamformer are jointly designed to maximize the radar receive SINR, where the MUI is designed to be constructive at the CUs.
\section{System Model}
We consider a dual-functional RadCom BS which serves $K$ single-antenna CUs and detects a point-like target in the presence of $I$ clutter sources simultaneously, and the BS is equipped with $N_T$ transmit antennas and $N_R$ receive antennas. The target is regarded as a potential eavesdropper which might intercept the information transmitted from BS to CUs.
Let ${\mathbf{x}} \in {\mathbb{C}^{{N_T} \times 1}}$ denote the transmit signal vector, the received waveform is given as
\begin{equation}\label{eq1}
    {\mathbf{r}} = {\alpha _0}{\mathbf{U}}\left( {{\theta _0}} \right){\mathbf{x}} + \sum\limits_{i = 1}^I {{\alpha _i}{\mathbf{U}}\left( {{\theta _i}} \right)} {\mathbf{x}} + {\mathbf{z}}
\end{equation}
where ${\alpha _0}$ and ${\alpha _i}$ denote the complex amplitudes of the target and the \emph{i}-th interference source, ${\theta _0}$ and ${\theta _i}$ are the angle of the target and the \emph{i}-th clutter source, respectively, and ${\mathbf{z}} \in {\mathbb{C}^{{N_R} \times 1}}$ is the additive white Gaussian noise (AWGN) vector, with the variance of $\sigma _R^2$. ${\mathbf{U}}\left( {{\theta}} \right)$ is the steering matrix of uniform linear array (ULA) antenna with half-wavelength spaced element, which is defined as
\begin{equation}\label{eq2}
    {\mathbf{U}}\left( \theta  \right) = {{{\mathbf{a}}_r}\left( \theta  \right){\mathbf{a}}_t^T\left( \theta  \right)}
\end{equation}
where ${{\mathbf{a}}_t}\left( \theta  \right) = \frac{1}{{\sqrt {{N_T}} }}{\left[ {1,{e^{ - j\pi \sin \theta }}, \cdots ,{e^{ - j\pi \left( {{N_T} - 1} \right)\sin \theta }}} \right]^T}$ and ${{\mathbf{a}}_r}\left( \theta  \right) = \frac{1}{{\sqrt {{N_R}} }}{\left[ {1,{e^{ - j\pi \sin \theta }}, \cdots ,{e^{ - j\pi \left( {{N_R} - 1} \right)\sin \theta }}} \right]^T}$. Then, the output of the filter can be given as
\begin{equation}\label{eq3}
\begin{split}
    {r_f} &= {{\mathbf{w}}^H}{\mathbf{r}}\\
          &= {\alpha _0}{{\mathbf{w}}^H}{\mathbf{U}}\left( {{\theta _0}} \right){\mathbf{x}} + \sum\limits_{i = 1}^I {{\alpha _i}{{\mathbf{w}}^H}{\mathbf{U}}\left( {{\theta _i}} \right)} {\mathbf{x}} + {{\mathbf{w}}^H}{\mathbf{z}},
\end{split}
\end{equation}
where ${\mathbf{w}}\in {\mathbb{C}^{{N_R} \times 1}}$ denotes the receive beamforming vector.
Accordingly, the output SINR can be expressed as
\begin{equation}\label{eq4}
\begin{split}
    {\text{SIN}}{{\text{R}}_{rad}} &= \frac{{{{\left| {{\alpha _0}{{\mathbf{w}}^H}{\mathbf{U}}\left( {{\theta _0}} \right){\mathbf{x}}} \right|}^2}}}{{{{\mathbf{w}}^H}\sum\limits_{i = 1}^I {{{\left| {{\alpha _i}} \right|}^2}{\mathbf{U}}\left( {{\theta _i}} \right){\mathbf{x}}{{\mathbf{x}}^H}{{\mathbf{U}}^H}\left( {{\theta _i}} \right){\mathbf{w}} + {{\mathbf{w}}^H}{\mathbf{w}}\sigma _R^2} }} \hfill \\
    {\text{           }}
    &= \frac{{\mu {{\left| {{{\mathbf{w}}^H}{\mathbf{U}}\left( {{\theta _0}} \right){\mathbf{x}}} \right|}^2}}}{{{{\mathbf{w}}^H}\left( {{\mathbf{\Sigma }}\left( {\mathbf{x}} \right) + {{\mathbf{I}}_{{N_R}}}} \right){\mathbf{w}}}} \hfill \\
\end{split}
\end{equation}
where $\mu  = {{ {{{\left| {{\alpha _0}} \right|}^2}} } \mathord{\left/
 {\vphantom {{\mathbb{E}\left[ {{{\left| {{\alpha _0}} \right|}^2}} \right]} {\sigma _R^2}}} \right.
 \kern-\nulldelimiterspace} {\sigma _R^2}}$, ${\mathbf{\Sigma }}\left( {\mathbf{x}} \right) = \sum\limits_{i = 1}^I {{b_i}{\mathbf{U}}\left( {{\theta _i}} \right){\mathbf{x}}{{\mathbf{x}}^H}{{\mathbf{U}}^H}\left( {{\theta _i}} \right)} $, and ${b_i} = {{ {{{\left| {{\alpha _i}} \right|}^2}} } \mathord{\left/
 {\vphantom {{\mathbb{E}\left[ {{{\left| {{\alpha _i}} \right|}^2}} \right]} {\sigma _R^2}}} \right.
 \kern-\nulldelimiterspace} {\sigma _R^2}}$.
\\\indent In the communication system, the received signal at the \emph{k}-th CU can be written as
\begin{equation}\label{eq5}
    {{y}_k} = {{\mathbf{h}}_k^H}{\mathbf{x}} + {{n}_k},
\end{equation}
where ${{\mathbf{h}}_k} \in {\mathbb{C}^{{N_T} \times 1}}$ denotes the multiple input single output (MISO) channel vector between the BS and the \emph{k}-th CU. Similarly, ${{n}_k}$ is the AWGN of the CU $k$ with the variance of $\sigma _{C_k}^2$. Additionally, we note that the intended symbol changed in symbol-level in the DM systems. Let $s_k$ denote the intended symbol of the $k$-th CU, which is $M$-PSK modulated. To this end, we define ${s_k} \in {\mathcal{A}_M}$, where ${\mathcal{A}_M} = \left\{ {{a_m} = {e^{j\left( {2m - 1} \right)\phi }},m = 1, \cdots ,M} \right\}$, $\phi  = {\pi  \mathord{\left/{\vphantom {\pi  M}} \right.\kern-\nulldelimiterspace} M}$, and $M$ denotes the modulation order.
\section{${\text{SIN}}{{\text{R}}_{rad}}$ Maximization With Known Target Location}
As demonstrated in \cite{kalantari2016directional}, the study of the DM technique can be based on strict phase and relaxed phase constraints. By the strict phase-based waveform design, the received signal $y_k$ should have exactly the same phase as the induced symbol of the $k$-th CU (i.e., $s_k$), which decreases the degrees of freedom (DoFs) in designing the waveform ${\mathbf{x}}$. Hence, inspired by the concept of CI \cite{masouros2015exploiting,masouros2009dynamic}, propose to restrict the received symbol for each CU within a constructive region rather than within a line,  namely the relaxed phase based design.
\\\indent The CI technique has been widely investigated in the recent work. To avoid deviating our focus, we will omit the derivation of the CI constraints, and refer the reader to \cite{masouros2015exploiting} for more details. Since CI-based waveform design aims to transform the undesirable MUI into useful power by pushing the received signal further away from the $M$-PSK decision boundaries, all interference contributes in the useful received power \cite{xu2020rethinking}. Herewith, the SNR of the $k$-th user is expressed as
\begin{equation}\label{eq6}
    {\text{SN}}{{\text{R}}_k} = \frac{{{{\left| {{\mathbf{h}}_k^H{\mathbf{x}}} \right|}^2}}}{{\sigma _{{C_k}}^2}}.
\end{equation}
\\\indent With the knowledge of the channel information, all CUs' data, as well as the location of target and clutter resources readily available at the transmitter, we formulate the following optimization problem aiming at maximizing the SINR of the target return
\begin{equation}\label{eq7}
\begin{gathered}
  \mathop {\max }\limits_{{\mathbf{w}},{\mathbf{x}}} \;\;\;\;{\text{  SIN}}{{\text{R}}_{rad}} \hfill \\
  s.t.\;\;\;\;{\text{    }}{\left\| {\mathbf{x}} \right\|^2} \leq {P_0} \hfill \\
  \;\;\;\;\;\;\;\; {\text{         }}\left| {\arg \left( {{\mathbf{h}}_k^H{\mathbf{x}}} \right) - \arg \left( {{s_k}} \right)} \right| \leq \xi  , \forall k, \hfill \\
  \;\;\;\;\;\;\;\;\;\;  {\text{SN}}{{\text{R}}_k} \geq {\Gamma _k},  \forall k, \hfill \\
\end{gathered}
\end{equation}
where $P_0$ denotes the transmit power budget, $\Gamma_k$ is the given SNR threshold, and $\xi$ is the phase threshold where the noise-less received symbols are supposed to lie.
\\\indent As illustrated in Fig. 1, by taking one of the QPSK constellation points as an example, the constructive region is given as the green area. In Fig. 1(a), ${{\bar y}_k}$ denotes $\left({{y_k} - {n_k}}\right)$ and the SNR related scalar $\gamma_k$ is the threshold distance to the decision region of the received symbol at the $k$-th CU. Then, in order to express the constructive region geometrically, we rotate the noise-free received signal ${{\bar y}_k}$ and project it into real and imaginary axes, which is illustrated in Fig. 1(b). By noting $\left| {{s_k}} \right| = 1$, the rotated signal $\tilde y_k$ can be given in the form of
\begin{equation}\label{eq8}
\begin{aligned}
    {{\tilde y}_k} = \left( {{y_k} - {n_k}} \right)\frac{{s_k^*}}{{\left| {{s_k}} \right|}}
    &= {\mathbf{h}}_k^H{\mathbf{x}}s_k^*\\ & = {\mathbf{\tilde h}}_k^H{\mathbf{x}},
\end{aligned}
\end{equation}
where ${{{\mathbf{\tilde h}}}_k} = {{\mathbf{h}}_k}s_k^*$. Let us represent $\operatorname{Re} \left( {{{\tilde y}_k}} \right) = \operatorname{Re} \left( {{\mathbf{\tilde h}}_k^H{\mathbf{x}}} \right)$ and $\operatorname{Im} \left( {{{\tilde y}_k}} \right) = \operatorname{Im} \left( {{\mathbf{\tilde h}}_k^H{\mathbf{x}}} \right)$. Then, the ${\text{SINR}_{rad}}$ maximization problem (\ref{eq7}) can be recast as \cite{masouros2015exploiting}
\begin{subequations}\label{eq9}
\begin{align}
  &\mathop {\max }\limits_{{\mathbf{w}},{\mathbf{x}}} {\text{  SIN}}{{\text{R}}_{rad}} \hfill \\
  &s.t.\;\;\;\;\; {\left\| {\mathbf{x}} \right\|^2} \leq {P_0} \hfill \\
  &\;\;\;\;\;\;\;\;\; \left| {\operatorname{Im} \left( {{\mathbf{\tilde h}}_k^H{\mathbf{x}}} \right)} \right| \leq \left( {\operatorname{Re} \left( {{\mathbf{\tilde h}}_k^H{\mathbf{x}}} \right) - \sqrt {\sigma _{{C_k}}^2{\Gamma _k}} } \right)\tan \phi, \forall k
\end{align}
\end{subequations}
where $\phi {\text{ = }} \pm {\pi  \mathord{\left/{\vphantom {\pi  M}} \right. \kern-\nulldelimiterspace} M}$.
\subsection{Solve (9) by Successive QCQP Approach}
It is noted that the problem (\ref{eq11}) is still difficult to solve due to the non-convex objective function. Firstly, by solving a well-known minimum variance distortionless response (MVDR) problem to maximize the output SINR in radar system, the corresponding receive beamforming vector can be expressed as the following function of ${\mathbf{x}}$
\begin{equation}\label{eq10}
    {\mathbf{w}} = \frac{{{{\left[ {{\mathbf{\Sigma }}\left( {\mathbf{x}} \right) + {\mathbf{I}}} \right]}^{ - 1}}{\mathbf{U}}\left( {{\theta _0}} \right){\mathbf{x}}}}{{{{\mathbf{x}}^H}{{\mathbf{U}}^H}\left( {{\theta _0}} \right){{\left[ {{\mathbf{\Sigma }}\left( {\mathbf{x}} \right) + {\mathbf{I}}} \right]}^{ - 1}}{\mathbf{U}}\left( {{\theta _0}} \right){\mathbf{x}}}}.
\end{equation}
By substituting (10) into (4), the optimization problem (\ref{eq11}) can be rewritten as \cite{7450660,cui2013mimo}
\begin{equation}\label{eq11}
\begin{aligned}
    &\mathop {\max }\limits_{{\mathbf{x}}} {\text{  }}{{\mathbf{x}}^H}{\mathbf{\Phi }}\left( {\mathbf{x}} \right){\mathbf{x}} \hfill \\
    &s.t.{\text{    }}9\left( b \right){\text{  }}{\text{and}}{\text{  }}9\left( c \right),
\end{aligned}
\end{equation}
where ${\mathbf{\Phi }}\left( {\mathbf{x}} \right)$ is a positive-semidefinite SINR matrix, which is expressed as ${\mathbf{\Phi }}\left( {\mathbf{x}} \right) = {\mathbf{U}}{\left( {{\theta _0}} \right)^H}{\left[ {{\mathbf{\Sigma }}\left( {\mathbf{x}} \right) + {\mathbf{I}}} \right]^{ - 1}}{\mathbf{U}}\left( {{\theta _0}} \right)$. To solve problem (\ref{eq11}), we adopt the sequential optimization algorithm (SOA) presented in \cite{cui2013mimo}. To be specific, let us firstly ignore the dependence of ${\mathbf{\Phi }}\left( {\mathbf{x}} \right)$ on $\mathbf{x}$, i.e., fixing the signal-dependent matrix ${\mathbf{\Phi }}\left( {\mathbf{x}} \right)={\mathbf{\Phi }}$ for a given ${\mathbf{x}}$. To start with, we initialize ${\mathbf{\Phi }}={\mathbf{\Phi }}_0$, where ${\mathbf{\Phi }}_0$ is a constant positive-semidefinite matrix. By using SOA, the waveform ${\mathbf{x}}$ is optimized iteratively with the updated ${\mathbf{\Phi}}$ till convergence. By doing so, in each SOA iteration we solve the following problem
\begin{equation}\label{eq12}
\begin{aligned}
    &\mathop {\max }\limits_{{\mathbf{x}}} {\text{  }}{{\mathbf{x}}^H}{\mathbf{\Phi }}{\mathbf{x}}\hfill \\
    &s.t.{\text{    }}9\left( b \right){\text{  }}{\text{and}}{\text{  }}9\left( c \right).
\end{aligned}
\end{equation}
Note that problem (12) is easily converted to a convex Quadratically Constrained
Quadratic Program (QCQP) problem by recasting the signal-independent matrix ${\mathbf{\Phi }}$ to be negative-semidefinite as follows \cite{7450660}
\begin{equation}\label{eq13}
\begin{aligned}
    \mathop {\max }\limits_{\mathbf{x}} &{\text{  }}{{\mathbf{x}}^H}{\mathbf{Qx}} \hfill \\
    s.t.{\text{    }}9&\left( b \right){\text{  }}{\text{and}}{\text{  }}9\left( c \right),
\end{aligned}
\end{equation}
where ${\mathbf{Q}} = \left( {{\mathbf{\Phi }} - \lambda {\mathbf{I}}} \right)$,  $\lambda  \ge {\lambda _{\max }}\left( {\mathbf{\Phi }} \right)$. It is straightforward to see that $\mathbf{Q}$ is negative-semidefinite, thus the objective function is concave, and then it can be tackled efficiently by CVX toolbox \cite{grant2014cvx}. Furthermore, as the expression given in (10), the receive beamforming vector $\mathbf{w^*}$ can be updated by the optimal waveform $\mathbf{x^*}$. Therefore, the suboptimal solutions are obtained until convergence by updating $\mathbf{x}$ and $\mathbf{w}$ iteratively. The generated solution will serve as a baseline in Section VI named as SQ.
\\\indent In SQ approach, we note that the reformulation of the objective function in (13) actually relaxes the one given in (12), to be specific, we have ${\mathbf{x}^{H}}{\mathbf{Qx}} = {\mathbf{x}^H}\left({{\mathbf{\Phi }} - \lambda {\mathbf{I}}} \right){\mathbf{x}} = {\mathbf{x}^{H}}{\mathbf{\Phi x}}-\lambda {\mathbf{x}^H}{\mathbf{x}}$, while the power constraint (9b) indicates that ${\mathbf{x}^H}{\mathbf{x}}$ in the second term is not constant. In the following subsection, we adopt SDR to solve problem (12), which aims to tackle the problem without a relaxation in the objection function.
\begin{figure}
    \centering
    \subfigure[]{
    \includegraphics[width=0.45\columnwidth]{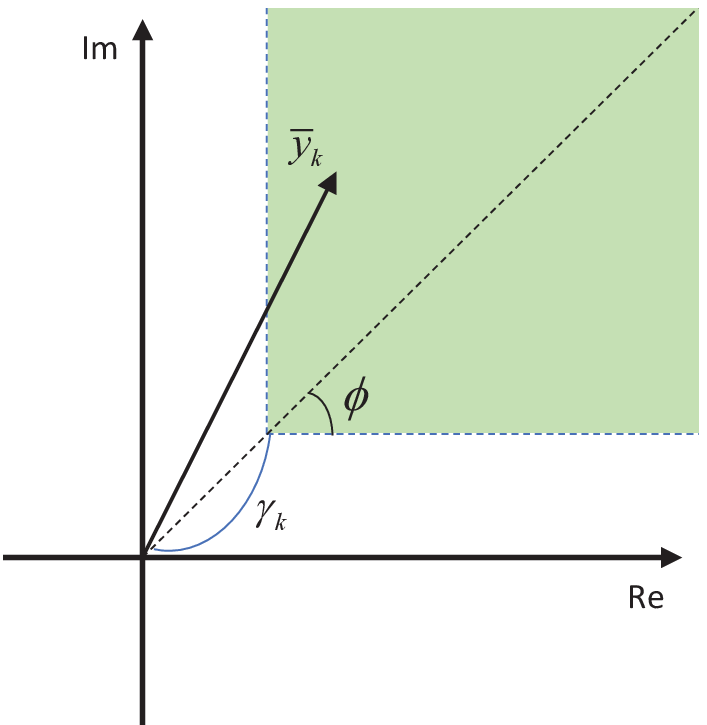}}
    \subfigure[]{
    \includegraphics[width=0.45\columnwidth]{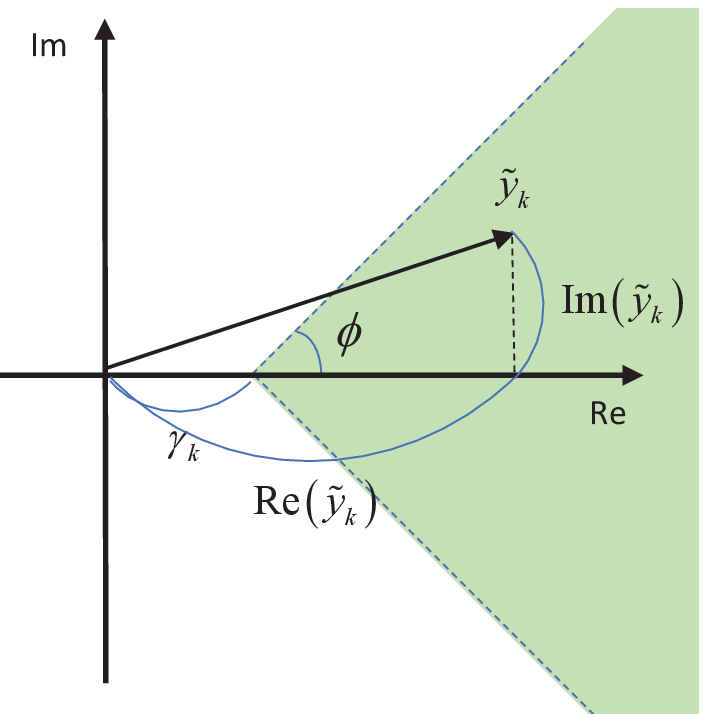}}
    \captionsetup{font={footnotesize}}
    \caption{QPSK illustration. (a) Relaxed phase DM. (b) Rotation by $\arg \left( {s_k^*} \right)$.}
    \label{fig.1}
\end{figure}
\subsection{Solve (9) by SDR Approach}
It is noted that problem (12) is an inhomogeneous QCQP \cite{vandenberghe1996semidefinite} problem. To tackle problem (12), we firstly define a new variable $\mathbf{X} = {\mathbf{x}}{\mathbf{x}}^H$ and let
\begin{equation}\label{eq14}
    {\mathbf{\tilde X}} = \left[ {\begin{array}{*{20}{c}}
    {\mathbf{X}}&{\mathbf{x}} \\
    {{{\mathbf{x}}^H}}&1
\end{array}} \right].
\end{equation}
\addtolength{\topmargin}{0.04in}
Afterwards, problem (14) can be recast as
\begin{equation}\label{eq15}
\begin{gathered}
  \mathop {\max }\limits_{\mathbf{X}, \mathbf{x}} {\text{  tr}}\left( {{\mathbf{X \Phi}}} \right) \hfill \\
  s.t.\;\;\;{\text{    tr}}\left( {\mathbf{X}} \right) \leq {P_0} \hfill \\
   \;\;\;\;\;\;\;\; \left| {\operatorname{Im} \left( {{\mathbf{\tilde h}}_k^H{\mathbf{x}}} \right)} \right| \leq \left( {\operatorname{Re} \left( {{\mathbf{\tilde h}}_k^H{\mathbf{x}}} \right) - \sqrt {\sigma _{{C_k}}^2{\Gamma _k}} } \right)\tan \phi {\text{ }} \hfill \\
   \;\;\;\;\;\;\;\;\;  {\mathbf{\tilde X}} \succeq 0, {\text{rank}}\left( {{\mathbf{\tilde X}}} \right) = 1. \hfill \\
\end{gathered}
\end{equation}
Note that problem (15) is readily to be solved by the SDR technique \cite{park2017general}. To start with, we relax the above optimization problem by dropping the rank-1 constraint, yielding
\begin{equation}\label{eq16}
\begin{gathered}
  \mathop {\max }\limits_{\mathbf{X}, \mathbf{x}} {\text{  tr}}\left( {{\mathbf{X \Phi}}} \right) \hfill \\
  s.t.\;\;\;{\text{    tr}}\left( {\mathbf{X}} \right) \leq {P_0} \hfill \\
   \;\;\;\;\;\;\;\; \left| {\operatorname{Im} \left( {{\mathbf{\tilde h}}_k^H{\mathbf{x}}} \right)} \right| \leq \left( {\operatorname{Re} \left( {{\mathbf{\tilde h}}_k^H{\mathbf{x}}} \right) - \sqrt {\sigma _{{C_k}}^2{\Gamma _k}} } \right)\tan \phi {\text{ }} \hfill \\
   \;\;\;\;\;\;\;\;\;  {\mathbf{\tilde X}} \succeq 0. \hfill \\
\end{gathered}
\end{equation}
Problem (16) is convex and can be optimally solved. We define ${{\mathbf{X}}^*}$ and ${{\mathbf{x}}^*}$ as the approximate solution to the problem above. By substituting the ${{\mathbf{X}}^*}$ in the objective function in (16), the optimal objective value is an upper bound of the optimal value in problem (9). If an optimal point of (16) satisfies ${{\mathbf{X}}^*} = {{\mathbf{x}}^*}{{\mathbf{x}}^{*H}}$, the SDR bound is tight, which indicates that ${{\mathbf{x}}^*}$ is a solution of problem (9). While the SDR bound is not tight in general, we adopt Gaussian randomization procedure \cite{luo2010semidefinite} to obtain the suboptimal solution to (9).
\subsection{Solve (9) by SCA Approach}
We note that the non-convexity lies only in the objective function in problem (12), and one can stay in the convex feasible region by exploiting various linear iteration schemes. Thus, it can be solved by converting the objective function into its linear approximation form. Inspired by the Frank-Wolfe approach \cite{frank1956algorithm}, we design a successive convex approximation (SCA) method to tackle problem (12). First of all, problem (12) can be equivalently transformed into
\begin{equation}\label{eq17}
\begin{aligned}
    \mathop {\min }\limits_{{\mathbf{x}}} {\text{  }}&f\left(x\right)=-{{\mathbf{x}}^H}{\mathbf{\Phi }}{\mathbf{x}}\hfill \\
    s.t.{\text{    }}&9\left( b \right){\text{  }}{\text{and}}{\text{  }}9\left( c \right).
\end{aligned}
\end{equation}
To proceed with the SCA technique, we approximate the objective function $f\left(x\right)$ by its first-order Taylor expansion with respect to $\mathbf{x}$ at ${\mathbf{x}}' \in \mathcal{D}$, where $\mathcal{D}$ denotes the feasible region of (17)
\begin{equation}\label{eq18}
\begin{aligned}
    f\left( {\mathbf{x}} \right) &\approx f\left( {{\mathbf{x}}'} \right){\text{ + }}\nabla f^H\left( {{\mathbf{x}}'} \right)\left( {{\mathbf{x}} - {\mathbf{x}}'} \right) \hfill \\
    & = f\left( {{\mathbf{x}}'} \right) + {\text{Re}}\left({\left( { - 2{\mathbf{\Phi x}}'} \right)^H}\left( {{\mathbf{x}} - {\mathbf{x}}'} \right)\right),
\end{aligned}
\end{equation}
where $\nabla f\left(  \cdot  \right)$ denotes the gradient of $f\left(  \cdot  \right)$. Herewith, the $m$-th iteration of the SCA algorithm can be obtained by solving the following convex optimization problem
\begin{equation}\label{eq19}
\begin{aligned}
    \mathop {\min }\limits_{\mathbf{x}} {\text{  }}g\left( {\mathbf{x}} \right) &=  {\text{Re}}\left(- 2{\left( {{\mathbf{\Phi x}}^{m-1}} \right)^H}\left({\mathbf{x}}-{\mathbf{x}}^{m-1} \right)\right)  \hfill \\
    s.t.{\text{    }}&9\left( b \right){\text{  }}{\text{and}}{\text{  }}9\left( c \right),
\end{aligned}
\end{equation}
where ${{\mathbf{x}}^{m-1}} \in \mathcal{D}$ is the point obtained at the $\left(m-1\right)$-th iteration. The optimal solution ${{\mathbf{x}}^ * } \in \mathcal{D}$ is generated by solving the problem (19). Since $g\left( {{{\mathbf{x}}^ {m-1} }} \right) = 0$, it is readily noted that $g\left( {{{\mathbf{x}}^ * }} \right) \leq 0$, which implies that ${{\mathbf{x}}^ * } - {{\mathbf{x}}^{m - 1}}$ is always a descent direction for the $m$-th iteration. Then, we move on to update the point ${{\mathbf{x}}^{m}}$ with a stepsize $t$ towards the descent direction, which is indicated as follows
\begin{equation}\label{eq20}
    {{\mathbf{x}}^{m}} = {{\mathbf{x}}^{m - 1}} + t\left( {{{\mathbf{x}}^*} - {{\mathbf{x}}^{m - 1}}} \right), t \in \left[ {0,1} \right].
\end{equation}
Since both ${{\mathbf{x}}^ * }$ and ${{\mathbf{x}}^{m - 1}}$ are drawn from $\mathcal{D}$, we have ${{\mathbf{x}}^{m}} \in \mathcal{D}$ due to the convexity of the feasible region. For clarity, the proposed SCA method to solve problem (12) is summarized in Algorithm 1. We note that the computational complexity of solving problem (19) at each iteration is given by $\mathcal{O} \left( N_T^3\sqrt {K + 1}  \right)$ \cite{nesterov1994interior}.
\begin{remark}\label{rmk:1}
    The performance of Algorithm 1 relies closely on the choice of initial point ${\mathbf{x}}^0$. To find a good one, the initial point can be generated by solving the following convex problem
    \begin{equation}\label{eq21}
    \begin{aligned}
        \mathop {\max }\limits_{\mathbf{x}} &{\text{  }}\sum\limits_{p=1}^{{N_t}} {{x_p}}   \hfill \\
        s.t.{\text{    }}9\left( b \right)&{\text{  }}{\text{and}}{\text{  }}9\left( c \right),
    \end{aligned}
    \end{equation}
    which aims to maximize the sum of elements in the vector $\mathbf{x}$ within the same feasible region.
\end{remark}
\renewcommand{\algorithmicrequire}{\textbf{Input:}}
\renewcommand{\algorithmicensure}{\textbf{Output:}}
\begin{algorithm}
\caption{The Proposed SCA Algorithm for solving problem (\ref{eq14})}
\label{alg:1}
\begin{algorithmic}
    \REQUIRE ${P_0},{\mathbf{h}}_k,{{\sigma _{{C_k}}^2}},{\sigma _R^2},\theta_i,\theta_0,\alpha_0, b_i, \Gamma_k, \forall k,\forall i, \varepsilon > 0$, and the maximum iteration number $m_{max}$
    \ENSURE ${\mathbf{x}}$
    \STATE 1. Reformulate the objective function by (17).
    \STATE 2. Initialize ${\mathbf{x}}^0 \in\ \mathcal{D}$ randomly, $m = 1$.
    \WHILE {$m \le {m_{max}}$ and $\left| g \left( {\mathbf{x}}^{m} \right) \right| \ge \varepsilon$ }
    \STATE 3. Set the gradient as $\nabla f\left( {{\mathbf{x}}}^{m-1} \right)$ and solve the problem (19) to obtain the optimal ${{\mathbf{x}}^*}$.
    \STATE 4. Update the ${{\mathbf{x}}^*}$ by (20), where $t$ is the stepsize which can be found by the exact line search or Armijo method.
    \STATE 5. $ m = m + 1$.
    \ENDWHILE
\end{algorithmic}
\end{algorithm}
\section{Numerical Results}
In this section, we evaluate the proposed methods via Monte Carlo based simulation results given as follows. Without loss of generality, each entry of the channel vector ${\mathbf{h}}_k$ is assumed to obey standard Complex Gaussian distribution. We assume that both the DFRC BS and the radar receiver are equipped with uniform linear arrays (ULAs) with the same number of elements with half-wavelength spacing between adjacent antennas. In the following simulations, the power budget is set as $P_0 = 30 \text{dBm}$. The target is located at $\theta_0={0^ \circ }$ with a reflecting power of ${{{\left| {{\alpha _0}} \right|}^2}} = 10 {\text{dB}}$ and clutter sources are located at $\theta_1={-50^ \circ }, \theta_2={-20^ \circ }, \theta_3={20^ \circ }, \theta_4={50^ \circ }$ reflecting a power of ${{{\left| {{\alpha _1}} \right|}^2}} = {{{\left| {{\alpha _2}} \right|}^2}} = {{{\left| {{\alpha _3}} \right|}^2}} = {{{\left| {{\alpha _4}} \right|}^2}} = 30 {\text{dB}}$.
\begin{figure}
    \centering
    \subfigure[]{
    \includegraphics[width=0.45\columnwidth]{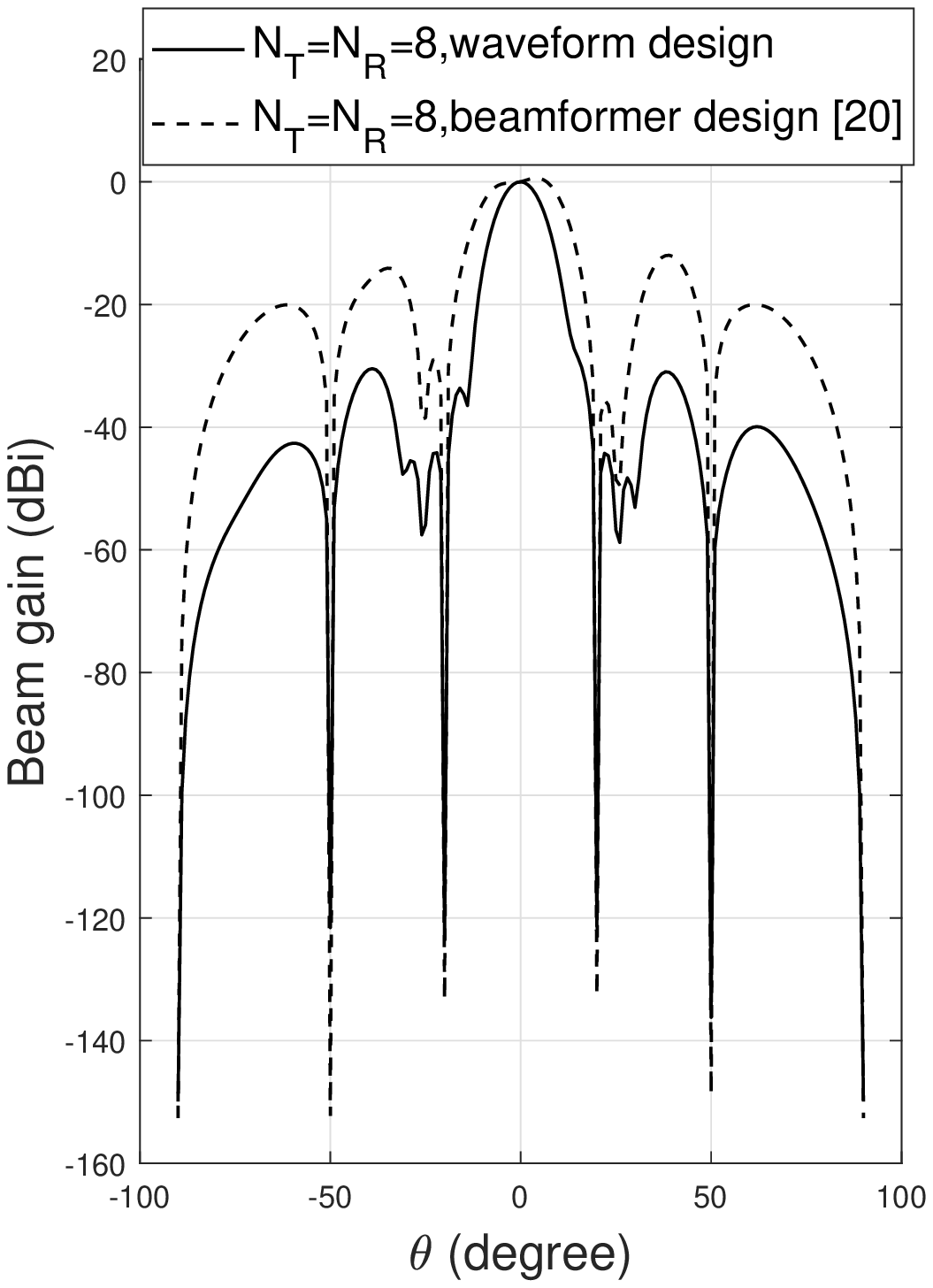}}
    \subfigure[]{
    \includegraphics[width=0.45\columnwidth]{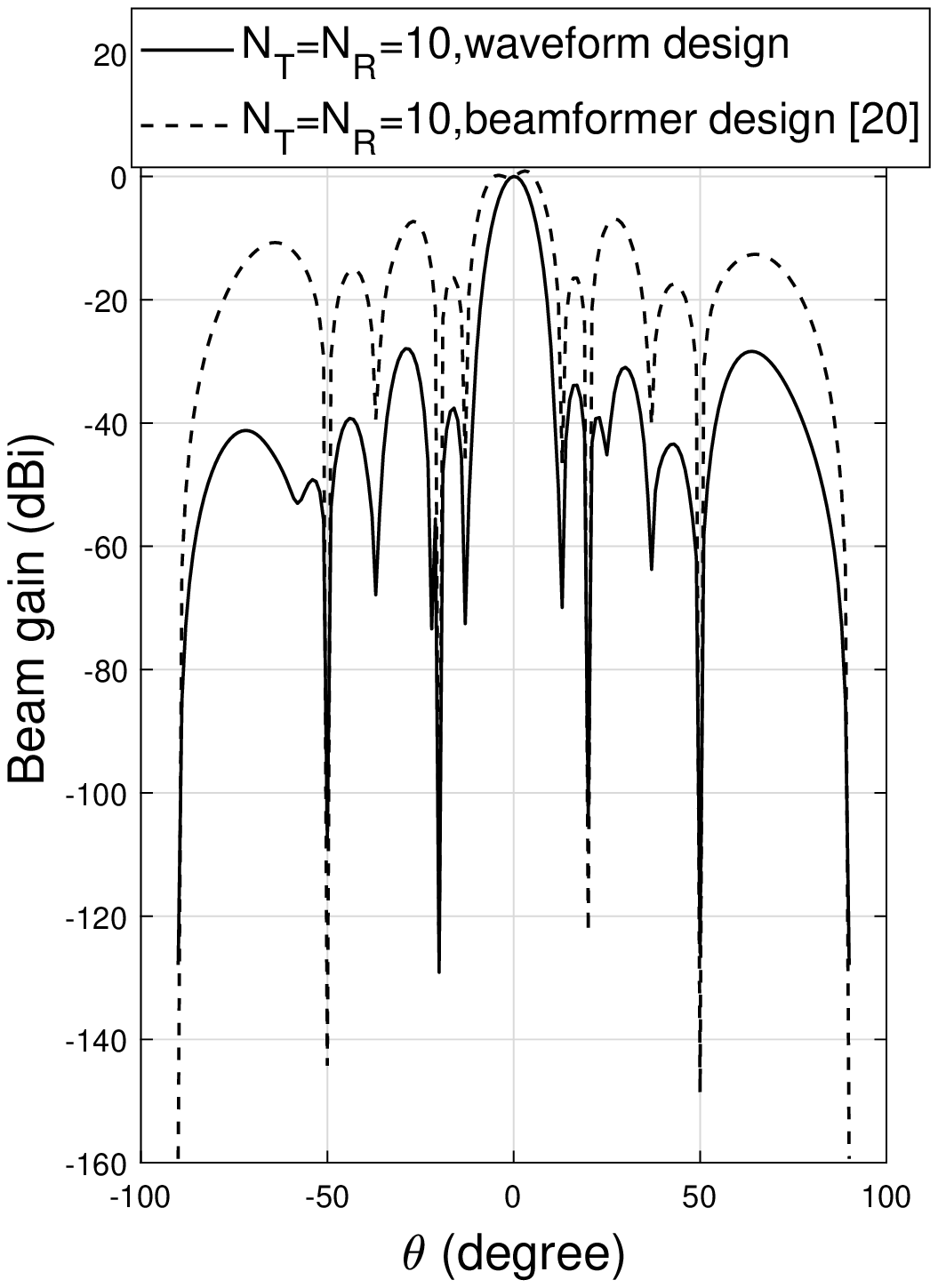}}
    \captionsetup{font={footnotesize}}
    \caption{Optimized beampatterns with different number of DFRC BS antennas, here, the beamformer design approach proposed in \cite{chen2020composite} is set as benchmarks, $K=5$.}
    \label{fig.2}
\end{figure}
\begin{figure}
    \centering
    \includegraphics[width=0.9\columnwidth]{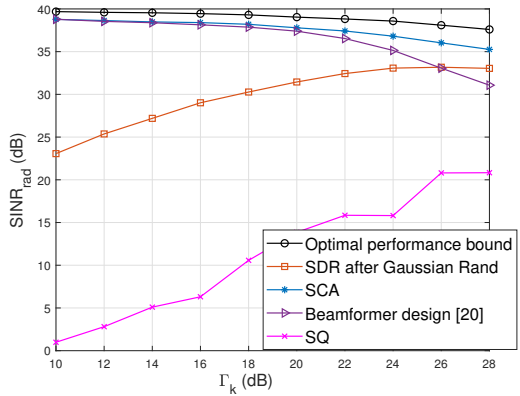}
    \captionsetup{font={footnotesize}}
    \caption{The performance of radar SINR versus CU's SNR with different solving methods, $N_T=N_R=8, K=5$.}
    \label{fig.3}
\end{figure}
\\\indent The resultant beampattern is firstly given in Fig. 2 with different number of DFRC BS antennas, where we set the presented beamformer design method in \cite{chen2020composite} as a benchmark. The SNR threshold $\Gamma_k, \forall k$ is fixed as 15dB. The nulls at the locations of clutter sources are clearly illustrated. It can be observed that the performance of beampattern gets better from the viewpoint of radar and the main beam width decreases with the increasing number of BS antennas. Additionally, comparing with the beamformer design method proposed in \cite{chen2020composite}, the peak to sidelobe ratio (PSLR) of the resultant beampattern generated from our proposed waveform design method is higher.
\\\indent The average performance of the tradeoff between the given SNR threshold of CU and the SINR of radar is illustrated in Fig. 3, including benchmark algorithms. In specific, with respect to the benchmarks, SQ denotes the proposed solving method in \cite{7450660}, SDR without Gaussian Rand denotes the optimal performance bound of the objective function as we have given in Section III, B. To satisfy the rank-1 constraint, Gaussian randomization procedure is commonly required, and the simulation result of which is given in Fig. 3 denoted as 'SDR after Gaussian Rand'. It is found that the received SINR of radar increases with the growth of $\Gamma_k$ when we adopt SQ method and the SDR technique after Gaussian randomization procedure, while $\text{SINR}_{rad}$ decreases when we deploy the other methods. This is for the reason that the optimized system power increases with the growth of $\Gamma_k$, which is less than the given power budget $P_0$, under the circumstance when SQ method or SDR solver with Gaussian randomization procedure is deployed. Moreover, the proposed waveform design method reaches a higher $\text{SINR}_{rad}$ comparing with the beamformer design in \cite{chen2020composite}, especially when $\Gamma_k$ is above 22dB.
\section{Conclusions}
In this paper, the transmit waveform and the receive beamformer have been jointly designed to maximize the received radar SINR for the DFRC system, objecting to fulfill the power budget and CI based security constraints, where the target location is assumed to be known at the BS perfectly. According to the numerical results, CI based waveform design has been indicated to outperform the beamformer design proposed in \cite{chen2020composite}, especially in the multi-CU scenario, the PSLR was higher in our design as shown in the beampattern. Also, the SCA solver has generated higher radar receive SINR comparing to SQ and SDR methods.
\section*{Acknowledgment}
This work has received funding from the Engineering and Physical Sciences Research Council (EPSRC) of the UK Grant number EP/S026622/1, the UK MOD University Defence Research Collaboration (UDRC) in Signal Processing and the China Scholarship Council (CSC).

\bibliographystyle{IEEEtran}
\bibliography{IEEEabrv,CEP_REF}

\end{document}